\documentclass[runningheads]{llncs}
\usepackage[T1]{fontenc}
\usepackage{graphicx}
\usepackage{amsmath}
\usepackage{booktabs}
\usepackage{diagbox}
\usepackage{subcaption}
\usepackage{bbding}
\begin{document}
\title{Modeling User Preferences as Distributions for Optimal Transport-Based Cross-Domain Recommendation under Non-Overlapping Settings}
\titlerunning{DUP-OT}
%
\author{Ziyin Xiao\inst{1}\orcidID{0009-0008-3327-6595}\Envelope \and
Toyotaro Suzumura\inst{2}\orcidID{0000-0001-6412-8386}
}
\authorrunning{Z. Xiao and T. Suzumura}
%
\institute{The University of Tokyo, Japan\\
\email{1014474276@qq.com}\and
The University of Tokyo, Japan\\
\email{suzumura@g.ecc.u-tokyo.ac.jp}}

%
\maketitle              
\begin{abstract}
Cross-domain recommender (CDR) systems aim to transfer knowledge from data-rich domains to data-sparse ones, alleviating sparsity and cold-start issues present in conventional single-domain recommenders. However, many CDR approaches rely on overlapping users or items to establish explicit cross-domain connections, which is unrealistic in practice. Moreover, most methods represent user preferences as fixed discrete vectors, limiting their ability to capture the fine-grained and multi-aspect nature of user interests.

To address these limitations, we propose DUP-OT (Distributional User Preferences with Optimal Transport), a novel framework for non-overlapping CDR. DUP-OT consists of three stages: (1) a shared preprocessing module that extracts review-based embeddings using a unified sentence encoder and autoencoder; (2) a user preference modeling module that represents each user's interests as a Gaussian Mixture Model (GMM) over item embeddings; and (3) an optimal-transport-based alignment module that matches Gaussian components across domains, enabling effective preference transfer for target-domain rating prediction.

Experiments on Amazon Review datasets show that DUP-OT outperforms single-domain baselines even without source-domain data, and achieves lower RMSE than the cross-domain baseline TDAR under strictly non-overlapping training settings, demonstrating its effectiveness in reducing large prediction errors for cold-start users. The implementation is available at \url{https://github.com/XiaoZY2000/dup-ot}.

\keywords{Cross-Domain Recommendation \and Non-overlapping Users \and Gaussian Mixture Model (GMM) \and Optimal Transport \and Cold-start Problem}
\end{abstract}

\section{Introduction}

With the rapid growth of online information, recommender systems have become essential for helping users navigate massive content \cite{R_Lü_2012}. However, conventional single-domain recommenders suffer from cold-start and data sparsity issues \cite{T_He_2024,L_Liu_2024,A_Zang_2023}. Cross-domain recommendation (CDR) alleviates these problems by transferring knowledge from a dense source domain to a sparse target domain \cite{A_Zhu_2018}. Yet most existing CDR methods rely on overlapping users or items in the training stage to bridge domains and typically represent user preferences as discrete vectors, which limits their ability to capture multi-aspect preferences.

In this work, we study the non-overlapping CDR scenario—where no users or items are shared across domains—and propose to represent user preferences using Gaussian Mixture Models (GMMs). This enables fine-grained, multi-aspect preference modeling, but also introduces two challenges: (1) domain discrepancy becomes harder to address without overlapping entities \cite{D_Guo_2024,C_Liu_2023}, and (2) GMM-based preference modeling significantly increases computation and is not directly compatible with standard optimal transport formulations.

Recent CDR approaches attempt to reduce domain discrepancy using contrastive learning, adversarial training, or distribution alignment \cite{L_Liu_2024,R_Wang_2020,C_Perera_2019}. Other studies explore distributional representations but either adopt simple Gaussian models or still rely on overlapping entities \cite{M_Zhu_2024}. Thus, how to build expressive GMM-based preference representations while enabling cross-domain alignment in a strictly non-overlapping setting remains an open problem.

To address this, we propose \textbf{DUP-OT} (Distributional User Preferences with Optimal Transport), a novel framework that models each user's preference as a GMM over item representations and aligns domains using optimal transport (OT). To reduce computational cost and make OT feasible, we assume that all users within a domain share a domain-level set of Gaussian components extracted from item representations, while each user learns personalized weights over these components. DUP-OT consists of three stages: (1) a shared preprocessing stage with a unified sentence encoder and autoencoder to obtain low-dimensional item and user embeddings; (2) a user preference modeling stage where each domain fits item-level GMM components and learns user-specific mixture weights; and (3) an OT-based alignment stage that transports GMM weights from the source to the target domain to enhance target-domain rating prediction.

Our contributions are summarized as follows:
\begin{itemize}
    \item We highlight the limitations of discrete vector representations in non-overlapping CDR and motivate the need for distributional preference modeling.
    \item We propose DUP-OT, a GMM-based framework that aligns cross-domain preferences using optimal transport on distributional representations.
    \item We validate DUP-OT on Amazon Review datasets, demonstrating improved RMSE over both single-domain and cross-domain baselines, with particular effectiveness in reducing large prediction errors for cold-start users.
\end{itemize}

\section{Related Work}

\subsubsection{Cross-Domain Recommendation.}
CDR aims to transfer knowledge from a rich source domain to a sparse target domain. Early approaches rely on overlapped users or items through representation alignment \cite{D_Li_2020}, adversarial learning \cite{R_Li_2022}, and graph-based propagation \cite{D_Guo_2021}. However, such overlaps are rarely available in practice.

Non-overlapping CDR methods typically learn shared latent spaces using embedding alignment \cite{L_Wang_2021}, generative modeling \cite{T_Salah_2021}, or distribution matching \cite{M_Zhu_2024}. While recent work explores Gaussian-based representations, they either use simple single-Gaussian models or still rely on partial overlaps. Modeling user preferences with expressive GMMs under fully non-overlapping settings remains underexplored.

\subsubsection{Optimal Transport in CDR.}
OT provides a principled tool for aligning distributions across domains. It has been applied to match user/item embeddings \cite{G_Li_2022} or latent distributions \cite{L_Liu_2024}. However, prior work does not integrate OT with GMM-based user preference modeling, nor address the computational challenges in fully non-overlapping scenarios.

\section{Preliminary}

\subsubsection{Definitions of Notations.} 
We consider two domains: the source domain $s$ and the target domain $t$.  
For each domain $x \in \{s, t\}$, we denote the user set and item set as $\mathcal{U}_x$ and $\mathcal{V}_x$.  
Although users or items may exist in both domains in reality, in our CDR setting we impose that
\emph{the model does not use any overlapping-user or overlapping-item information during training}.  
Therefore, the training process treats the two domains as non-overlapping, even if overlaps may exist in the raw data.

For the $i$-th user and $j$-th item in domain $x$, each interaction is written as:
\[
(u_i^x,\, v_j^x,\, r_{ij}^x,\, h_{ij}^x,\, t_{ij}^x),
\]
where $r_{ij}^x$ is the rating, $h_{ij}^x$ is the corresponding review text and $t_{ij}^x$ is the time stamp.


\subsubsection{Practical Motivation.}  
This setting corresponds to real-world scenarios where user or item identities might be shared across platforms,  
but privacy constraints or system limitations prevent such cross-domain links from being used during training.  
For example, users may later appear in the target domain after the model has been trained,  
or items may be synchronized across platforms only at deployment time.  
Thus, overlaps can exist for evaluation but are not exploited for learning,  
faithfully reflecting practical cross-domain recommendation conditions.

\subsubsection{Task Formulation.} We aim to address the cold-start problem in the target domain by transferring knowledge from the source domain under a non-overlapping CDR setting. Specifically, the goal is to leverage source-domain data to enhance the rating prediction performance in the target domain.

\section{Methodology}
\subsection{Overview}
\begin{figure*}[t]
    \centering
    \hspace*{-15mm}
    \includegraphics[width=1.35\linewidth]{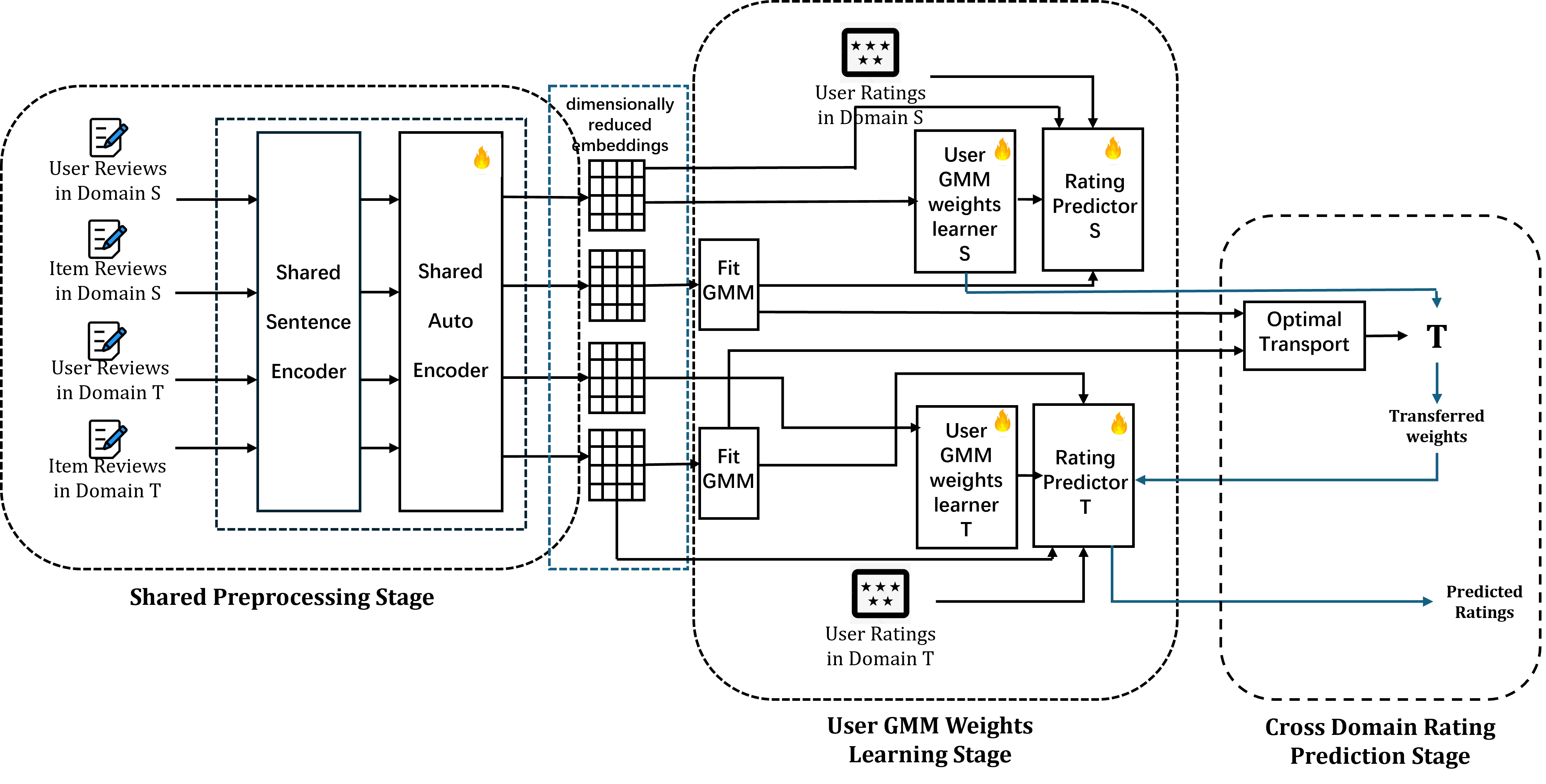}
    \caption{The overall structure of DUP-OT}
    \label{overall_structure}
\end{figure*}

Figure \ref{overall_structure} illustrates the architecture of our proposed \textbf{DUP-OT} framework, which enables non-overlapping cross-domain recommendation based on distributional user preference representations. The pipeline consists of three main stages: \textbf{(1) shared preprocessing}, \textbf{(2) user GMM weight learning}, and \textbf{(3) cross-domain rating prediction}.

In the \textbf{shared preprocessing stage}, we extract and align user and item embeddings across domains. Specifically, we employ a shared pre-trained sentence encoder to encode reviews into initial user and item embeddings, followed by a shared autoencoder trained on both domains to reduce embedding dimensionality and ensure a unified feature space.

The \textbf{user GMM weight learning stage} is performed separately in each domain. We first fit a Gaussian Mixture Model (GMM) on the item embeddings to obtain shared Gaussian components. Then, an MLP is trained to map each user embedding to a set of weights over these components, forming a user-specific GMM. Another MLP is trained to predict user-item ratings based on the Mahalanobis distance between the item embedding and each Gaussian component weighted by the user’s distribution.

Finally, in the \textbf{cross-domain rating prediction stage}, we align the GMM components across domains via Optimal Transport (OT) and use the resulting transport matrix to transfer users’ GMM weights from the source domain to the target domain. For users in the target-domain test set, our framework handles three cases: if a user has interactions in both domains during training, we combine the transferred distribution with the target-domain distribution; if the user has interactions only in the source domain, we use the transferred distribution; and if the user has interactions only in the target domain, we use the target-domain distribution. These resulting user distributions are then used to enhance rating prediction performance in the target domain.

\subsection{Shared Preprocessing Stage}

To capture semantic similarity across domains, we derive user and item embeddings from review text. For each user and item, all associated reviews are encoded using a shared pre-trained sentence encoder, and we apply a time-aware weighting scheme so that recent reviews contribute more to the final representation. Specifically, more recent interactions are assigned higher weights when aggregating review embeddings.

The encoder outputs high-dimensional vectors, which make GMM fitting unstable. Therefore, we train a shared autoencoder on both domains to obtain compact, domain-consistent embeddings. This unified latent space preserves semantic structure while reducing dimensionality, providing stable inputs for subsequent GMM construction and cross-domain alignment.

\subsection{User GMM Weights Learning Stage}

Users often exhibit multi-aspect preferences, which discrete vector representations struggle to capture. Therefore, DUP-OT models each user's preference as a Gaussian Mixture Model (GMM) over item embeddings.

To reduce computational cost, we assume that all users within a domain share a domain-level set of Gaussian components. These components are obtained by fitting a GMM on item embeddings using the EM algorithm (implemented via \texttt{BayesianGaussianMixture}). Each user is then mapped to a personalized set of mixture weights over these components.

For each domain, we train two MLPs: a weight learner that predicts user-specific mixture weights from user embeddings, and a rating predictor that takes the weighted component-wise Mahalanobis distances between a user and an item to estimate the final rating. The learning objective is the standard mean squared error:

\begin{align}
    \mathbf{w}^x_{u_i} &= \text{W-Learner}^x(\mathbf{z}^x_{u_i}),\\
    \hat{r}^x_{ij} &= \text{R-Predictor}^x\big(\mathbf{w}^x_{u_i},\, \mathbf{z}^x_{v_j},\, \mu^x,\, \Sigma^x\big),\\
    \mathcal{L}_{\mathrm{rating}}^x &= \frac{1}{|\mathcal{I}^x|} \sum_{(i,j) \in \mathcal{I}^x}\!\!\left(r_{ij}-\hat{r}_{ij}^x\right)^2.
\end{align}

Here, $\mu^x$ and $\Sigma^x$ denote the domain-level Gaussian component means and covariances. This stage produces a distributional user representation that is later aligned across domains via optimal transport.

\subsection{Cross-Domain Rating Prediction Stage}
In this stage, we enhance the target-domain rating predictor by constructing improved user distributions through Optimal Transport (OT). This stage consists of two parts: aligning the GMM components across domains using OT, and generating enhanced user-specific distributions in the target domain according to three test-time scenarios.

\subsubsection{OT-Based Alignment of Gaussian Components.}
Since all users within a domain share the same set of Gaussian components, we apply OT only between the two domains' GMM component sets, avoiding instance-level transport. The OT problem is formulated as Eq. \ref{eq:OT}.

\begin{equation}
\begin{aligned}
\text{arg}\min_{\mathcal{T}} \quad & \sum_{i=1}^{m} \sum_{j=1}^{n} \mathcal{T}_{ij} \mathcal{C}_{ij} \\
\text{s.t.} \quad & \sum_{i=1}^{m} \mathcal{T}_{ij} = \frac{1}{n}, \quad \sum_{j=1}^{n} \mathcal{T}_{ij} = \frac{1}{m}
\end{aligned}
\label{eq:OT}
\end{equation}

where $m$ and $n$ denote the numbers of Gaussian components in the source and target domains.  
The cost matrix $\mathcal{C}$ is computed using the Wasserstein-2 distance (Eq.~\ref{eq:w2-gaussian}), and we solve for the optimal transport matrix $\mathcal{T}$ using the Sinkhorn algorithm.

\begin{align}
\mathcal{W}_2^2(\mathcal{N}_1, \mathcal{N}_2) =\;
& \left\| \boldsymbol{\mu}_1 - \boldsymbol{\mu}_2 \right\|_2^2 \nonumber \\
& + \operatorname{Tr} \left( \boldsymbol{\Sigma}_1 + \boldsymbol{\Sigma}_2 
- 2\left( \boldsymbol{\Sigma}_2^{1/2} \boldsymbol{\Sigma}_1 \boldsymbol{\Sigma}_2^{1/2} \right)^{1/2} \right)
\label{eq:w2-gaussian}
\end{align}

User weights are then transported as Eq. \ref{eq:ot_u}.

\begin{equation}
    \mathbf{w}^t_{u_i}=\mathbf{w}^s_{u_i}\mathcal{T}
\label{eq:ot_u}
\end{equation}

\subsubsection{Three Test-Time Cases.}
For each user in the target-domain test set, we construct a final user distribution in the target domain depending on the interaction availability:
\begin{itemize}
    \item \textbf{Both domains have interactions:} we linearly fuse the transferred distribution and the target-domain distribution to form the final user distribution:
    \[
    \mathbf{w}_{u_i}^{\text{final}} = \alpha\, \mathbf{w}_{u_i}^{t} + (1-\alpha)\, \mathbf{w}_{u_i}^{t,\text{orig}},
    \]
    where $\alpha \in [0,1]$ controls the contribution of the transferred distribution.
    
    \item \textbf{Only source domain has interactions:} use the transferred distribution alone:
    \[
    \mathbf{w}_{u_i}^{\text{final}} = \mathbf{w}_{u_i}^{t}.
    \]

    \item \textbf{Only target domain has interactions:} use the target-domain distribution alone:
    \[
    \mathbf{w}_{u_i}^{\text{final}} = \mathbf{w}_{u_i}^{t,\text{orig}}.
    \]
\end{itemize}

For users without interactions in either domain, we initialize their mixture weights with random values. The resulting user distribution is then fed into the target-domain rating predictor to produce the final enhanced rating prediction.

Since all users within a domain share the same set of Gaussian components and only learn personalized mixture weights rather than full GMM parameters, the per-user computation is comparable to standard MLP-based models. Moreover, OT alignment operates only on the $m \times n$ component-level cost matrix (where $m, n \ll |\mathcal{U}|$), so the transport step adds negligible overhead compared to user-level MLP training.

\section{Experiments and Results}
We evaluate DUP-OT against several baselines to answer the following research questions:

\textbf{RQ1.} Does introducing cross-domain information enhance the recommendation performance in the target domain?

\textbf{RQ2.} Does modeling user preferences as distributions, rather than vectors, lead to improved recommendation performance?

\textbf{RQ3.} How does the performance of our proposed method compare with that of existing CDR models?

To answer \textbf{RQ1}, we conduct an ablation study comparing target-domain performance with and without source-domain information.
To address \textbf{RQ2}, we evaluate our target-domain model (using distribution-based user preferences but without source-domain data) against single-domain baselines.
To address \textbf{RQ3}, we compare our proposed DUP-OT model with a representative cross-domain recommendation baseline: TDAR.

\subsection{Experimental Setup}


\begin{table*}[t]
\centering
\caption{Results of baseline models and DUP-OT}
\label{tab:results}

\begin{subtable}{\textwidth}
\centering
\caption{Results on the Domain Pair $(D_s, D_t)$}
\hspace*{-17mm}
\begin{tabular}{l cc cc cc}
\toprule
$D_s \to D_t$
& \multicolumn{2}{c}{Digital Music $\to$ Electronics}
& \multicolumn{2}{c}{Movies \& TV $\to$ Electronics}
& \multicolumn{2}{c}{Video Games $\to$ Electronics} \\
\cmidrule(lr){2-3}
\cmidrule(lr){4-5}
\cmidrule(lr){6-7}
Model & RMSE & MAE & RMSE & MAE & RMSE & MAE \\
\midrule
TDAR 
& 1.2512 & \textbf{0.8349} 
& 1.2550 & \textbf{0.8378}
& 1.2484 & \textbf{0.8240} \\

DUP-OT (w/o source)
& 1.4632 & 1.0658
& 1.4055 & 0.9734
& 1.4407 & 1.0354 \\

DUP-OT (w/ source)
& \textbf{1.2283} & 0.9315
& \textbf{1.2499} & 1.0332
& \textbf{1.2385} & 0.9815 \\
\bottomrule
\end{tabular}
\end{subtable}

\vspace{1em}

\begin{subtable}{\textwidth}
\centering
\caption{Single-domain recommendation results (Electronics-only)}
\begin{tabular}{l cc}
\toprule
Model & RMSE & MAE \\
\midrule
LightGCN & 1.5317 & 1.1179 \\
NeuMF    & 1.4599 & 1.3297 \\
\midrule
DUP-OT (w/o source average)  
& \textbf{1.4365} & \textbf{1.0249} \\  
\bottomrule
\end{tabular}
\end{subtable}

\end{table*}

\subsubsection{Datasets}
We use the Amazon Review 5-core datasets \cite{J_Ni_2019}: Digital Music, Movies and TV, Video Games, and Electronics. Electronics is used as the target domain, and the other three as source domains. Target-domain data are split into train/validation/test (8:1:1) chronologically. To keep temporal consistency, we filter out source-domain interactions occurring later than the earliest validation timestamp in the target domain. The filtered source data and target-domain training set are used to train DUP-OT, and evaluation is performed on the target-domain test set.

\subsubsection{Model Setting}
Our proposed DUP-OT consists of three stages. In the first stage, we use a shared pre-trained sentence encoder (\texttt{all-MiniLM-L6-v2}) from Sentence-Transformers to obtain review-based user and item embeddings, followed by a shared autoencoder to reduce dimensionality and form a unified latent space across domains. In the second stage, for each domain we fit item-level Gaussian Mixture Models (GMMs) and train two small MLPs: one to predict user-specific mixture weights over the shared Gaussian components, and one to predict ratings based on the resulting user distributions and item embeddings. The number of GMM components $K$ is determined automatically by the \texttt{BayesianGaussianMixture} estimator.

\subsubsection{Training Settings}
We train all models using the Adam optimizer with a learning rate of \(3\times10^{-4}\) and a batch size of 256, with early stopping based on target-domain validation loss. The sentence encoder is frozen, while the shared autoencoder is jointly trained across domains to obtain unified low-dimensional embeddings. For each domain, we fit item-level Gaussian Mixture Models using the \texttt{BayesianGaussianMixture} estimator and train domain-specific MLPs for predicting user GMM weights and ratings. Users appearing only in validation or test sets are treated as cold-start users. All models are implemented in PyTorch and trained on a single NVIDIA GPU.

\subsubsection{Baselines} We implement the following models as baseline models for comparison:

\textbf{Single-Domain Models:}

\textbf{$\bullet$LightGCN\cite{L_He_2020}:} LightGCN is a simplified GCN model. In our experiment, we see it as a single domain RS model and build graphs on each domain rather than building a unified graph.

\textbf{$\bullet$NeuMF\cite{10.1145/3038912.3052569}:} A neural collaborative filtering model that fuses a generalized matrix factorization branch with a multi‑layer perceptron branch to jointly capture both linear and non‑linear user–item interaction patterns.

\textbf{Non-Overlapping Cross-Domain Models:}

\textbf{$\bullet$TDAR\cite{10.1145/3394486.3403264}:}
A text-enhanced domain-adaptation model that integrates textual and collaborative embeddings and uses adversarial learning for cross-domain alignment. Since our datasets contain explicit ratings, we replace its original contrastive objective with a rating-prediction loss.


For single domain RS models, we utilize RecBole \cite{recbole[1.0],recbole[2.0],recbole[1.2.1]} in our experiments to implement them. For TDAR, we implemented the method ourselves based on the original paper.

\subsubsection{Ablation Study: Effect of Source-Domain Information}

To evaluate whether source-domain information enhances recommendation performance in the target domain (RQ1), we compare two variants of our framework:

\begin{itemize}
    \item \textbf{DUP-OT (w/ source)}: the full model that uses both domains. The autoencoder is trained jointly on both domains, and in the second stage the source-domain GMMs, GMM-weight learners, and rating predictors are trained using source-domain data. User distributions are then transferred to the target domain via Optimal Transport and fused with target-domain distributions during inference.

    \item \textbf{DUP-OT (w/o source)}: an ablated model that does \emph{not} use any source-domain interactions during the second stage. The autoencoder is still trained jointly on both domains to ensure consistent representation spaces, but the GMM construction, GMM-weight learner, and rating predictor are trained exclusively on the target-domain training set. No cross-domain alignment or distribution transfer is performed.
\end{itemize}

This setup ensures that the only difference between the two variants lies in whether cross-domain preference information is introduced.

\subsubsection{Metrics} To evaluate the performance of our proposed DUP-OT and baseline models, we set all models' targets as rating prediction on the target domain's test dataset. And then we use RMSE and MAE as metrics to evaluate models' performance. The results are shown in Table \ref{tab:results}.

\subsection{Discussion of Results}

Based on the results in Table~\ref{tab:results}, we now answer the three research questions raised at the beginning of this section.

\subsubsection{Answer to RQ1 (Effect of Cross-Domain Information).}  
Across all three source$\to$target settings, DUP-OT (w/ source) consistently achieves lower RMSE than DUP-OT (w/o source). The effect on MAE is mixed, with improvements in some settings but not others, likely due to varying domain gaps between different source domains and the target domain. Nevertheless, the consistent RMSE reduction confirms that cross-domain preference transfer effectively mitigates large prediction errors, which is the primary goal of DUP-OT in cold-start scenarios.

\subsubsection{Answer to RQ2 (Effect of Distribution-Based User Modeling).}  
Comparing DUP-OT(w/o source) with single-domain recommenders (LightGCN and NeuMF), DUP-OT(w/o source) achieves substantially better performance in the Electronics-only setting, despite using no source-domain data. This indicates that modeling user preferences as distributions rather than discrete vectors yields more expressive representations and improves rating prediction accuracy in the target domain.

\subsubsection{Answer to RQ3 (Comparison with the Cross-Domain Baseline).} Compared with TDAR, DUP-OT (w/ source) achieves lower RMSE but higher MAE across all dataset combinations. RMSE is more sensitive to large errors, while MAE treats all errors equally. The lower RMSE indicates that DUP-OT effectively avoids extreme mispredictions, which are common for cold-start users with little interaction history. This can be explained by our GMM-based preference modeling: distributional representations provide more robust estimates for data-sparse users than point vectors, reducing the risk of severe prediction failures. On the other hand, TDAR's joint cross-domain optimization tends to produce more evenly spread errors, leading to lower MAE but less protection against worst-case predictions. This trade-off aligns well with the cold-start scenario that DUP-OT targets.

\section{Conclusion}

We proposed DUP-OT, a distribution-based framework for non-overlapping cross-domain recommendation. By modeling user preferences as GMMs and aligning domains via Optimal Transport, DUP-OT enables expressive preference representation and effective cross-domain transfer without relying on shared users or items during training. Experiments on Amazon Review datasets show that DUP-OT outperforms single-domain baselines even without source-domain data, and achieves lower RMSE than the cross-domain baseline TDAR, indicating stronger robustness against large prediction errors in cold-start scenarios.

A current limitation is that the linear fusion of transferred and target-domain distributions may not be optimal for all users. Future work includes exploring adaptive fusion strategies, extending DUP-OT to implicit feedback settings, and incorporating structure-aware transport costs such as Gromov-Wasserstein distance for better cross-domain alignment.

\paragraph{Disclosure of Interests.}
The authors have no competing interests to declare.

\bibliographystyle{splncs04}
\bibliography{reference}
\end{document}